\documentstyle[12pt]{article}
\author{Hans - J\"urgen Schmidt}
\title{Comment to the paper ``Conformal transformations single
out a  unique measure of distance'' Phys. Rev. D { \bf 47}, 1437    
(1993) by S. Cotsakis
}
\date{}
\begin{document}
\maketitle

\bigskip

\centerline{
Universit\"at Potsdam, Institut f\"ur Mathematik, Projektgruppe 
Kosmologie}
\centerline{
      D-14415 POTSDAM, PF 601553, Am Neuen Palais 10, Germany}

\bigskip

\begin{abstract} 
We discuss the consequences of the incorrectness [see the Erratum
in Phys. Rev D {\bf 49}, 1145 (1994)] of that paper and add two
related remarks. The scope of this comment is to encourage 
 further research on:  "Which of the conformally equivalent
metrics is the physical one?"
\end{abstract}


{\large
Cotsakis [1] considers the D-dimensional Lagrangian density 
\begin{equation}
L _g = f(R) \sqrt{-g}
\end{equation}
where $ R $ is the Ricci scalar. (One needs $ D > 2 $ to ensure 
the validity of the following formulas.) With inclusion of matter 
one gets from eq.(1) the following field equation (which 
coincides
with [1, eq.(6)])
\begin{equation}
h R \sp{ab}   - \frac{1}{2} f g \sp{ab}  
- \nabla \sp a  \nabla \sp b h + g \sp{ab} \Box h
= T \sp{ab} _M(g)  
\end{equation}
where $ h = df/dR $.
Eq. (1) is a scalar density, so the divergence of its variational

derivative with respect to the metric (i.e., of the l.h.s. of 
eq.(2) ) automatically vanishes. So this takes place for the 
r.h.s. also, i.e. (cf. the erratum to [1])
\begin{equation}
\nabla _a  T \sp{ab} _M(g) = 0
\end{equation}
whereas the original paper [1, eq.(21)] contains the superfluous 
expression
\begin{equation}
R \sp{ab} \nabla _a h
\end{equation}
This has as consequence that the  
 "class of manifolds" mentioned at the end of p.
1438 is empty; so Cotsakis' idea to decide between the different
metrics by a concrete model sounds interesting, but his special
example does not suffice to decide this, i.e., the question about
which is the correct "physical" metric remains to be answered  
yet. 

     Let us add two remarks to clarify the discussion. First, 
for
calculating the divergence of the l.h.s. of eq.(2) one must 
notice
that for an arbitrary scalar $ \psi $, the expressions 
$ \nabla \sp a \nabla \sp b \nabla \sp c \psi $
and $ \nabla \sp b \nabla \sp a \nabla \sp c \psi $
do not coincide. Their difference is the product of the Riemann 
tensor with  $  \nabla \sp d \psi $. Applying this with $ \psi = 
h $, one gets an expression which cancels the superfluous 
expression (4).

 Second,  the question which of the two metrics is the physical
one, 
was already discussed in several papers, e.g. ref.[2].
It holds { \bf A: } In the vicinity of flat space-time
and  also in the vicinity of the inflationary de Sitter
space-time
the  conformal factor is approximately a constant, so that
geodesic 
motion in the two metrics is almost the same. { \bf B: } Far away 
from these regions, in very strong fields, it is even not clear, 
whether geodesic motion takes place at all. 

\bigskip

{\it Acknowledgement}. I thank Dr. U. Kasper for independently
checking  the arguments; he also observed that Cotsakis
incorrectly 
applied Bochner's theorem, because that theorem refers to a 
positive definite metric only, one should try to generalize it to
indefinite metrics. 

\bigskip

{\it References}

[1] S. Cotsakis, Phys. Rev.   D  {\bf   47}, 1437 (1993), 
 D {\bf 49}, 1145(E) (1994).

[2] K. Maeda, J. Stein-Schabes, T. Futamase, Phys. Rev.  D {\bf   
39}, 2848 (1989). 
}

{\small{published in Phys. Rev. D {\bf     52}, 6198 (1995).
 Received 27 August 1993, revised 30 June 1994.} }

\end{document}